\documentclass[aps,prl,twocolumn,superscriptaddress,showpacs]{revtex4}
\usepackage{graphicx}
\begin{document}

\title{A one dimensional hard-point gas as a thermoelectric engine}

\author{Jiao Wang}
\affiliation{Temasek Laboratories, National University of
Singapore, 117542 Singapore} \affiliation{Beijing-Hong
Kong-Singapore Joint Center for Nonlinear and Complex Systems
(Singapore), National University of Singapore, 117542 Singapore}
\author{Giulio Casati}
\email[]{giulio.casati@uninsubria.it} \affiliation{Center for
Nonlinear and Complex Systems, Universit\`a degli Studi
dell'Insubria, Como, Italy} \affiliation{CNR-INFM and Istituto
Nazionale di Fisica Nucleare, Sezione di Milano, Milan, Italy }
\affiliation{ Centre for Quantum Technologies, National University
of Singapore, Singapore 117543}
\author{Toma\v z Prosen}
\affiliation{Physics department, Faculty of Mathematics and Physics,
University of Ljubljana, Ljubljana, Slovenia}
\author{ C.-H. Lai}
\email[]{cqtlch@nus.edu.sg} \affiliation{Beijing-Hong
Kong-Singapore Joint Center for Nonlinear and Complex Systems
(Singapore), National University of Singapore, 117542 Singapore}
\affiliation{ Centre for Quantum Technologies, National University
of Singapore, Singapore 117543}
\date{\today}

\begin{abstract}
We demonstrate the possibility to build a thermoelectric engine
using a one dimensional gas of molecules with unequal masses and
hard-point interaction. Most importantly, we show that the
efficiency of this engine is determined by a new parameter $YT$
which is different from the well known figure of merit $ZT$. Even
though the efficiency of this particular model is low, our results
shed new light on the problem and open the possibility to build
efficient thermoelectric engines.
\end{abstract}

\pacs{74.25.Fy,84.60.Rb,05.60.Cd,05.45.Pq} \maketitle

A combination of mathematical results, numerical studies and even
laboratory experiments have greatly improved our understanding of
phenomenological transport equations in recent years. The derivation
of such equations from purely dynamical laws, classical or quantum,
has been one of the main subject of interest \cite{rev}. Even though
a complete rigorous picture is still lacking, it is however clear
that dynamical chaos is an essential ingredient.

A better understanding of the above problem is important not only
from a fundamental point of view in order to provide a
justification of phenomenological laws. It is also relevant for
several applications. One particular important aspect is the
connection with thermoelectric power generation and refrigeration
\cite{mahan,majumdar,dresselhaus}. Indeed, due to the increasing
environmental concern and energy demand, thermoelectric phenomena
are expected to play an increasingly important role in meeting the
energy challenge of the future. However, the main difficulty is
the poor efficiency of existing devices. Indeed the suitability of
a thermoelectric material for energy conversion or electronic
refrigeration is characterized by the thermoelectric figure of
merit $Z=\sigma S^2/\kappa$,
%\begin{equation}
%\label{eq:ZT-def} Z = \frac{\sigma S^2}{\kappa}
%\end{equation}
where $\sigma$ is the coefficient of electric conductivity, $S$ is
the Seebeck coefficient and $\kappa$ is the thermal conductivity.
The Seebeck coefficient $S$, also called thermopower, is a measure
of the magnitude of an induced thermoelectric voltage in response
to a temperature difference.

For a given material, and a pair of temperatures $T_H$ and $T_C$
of hot and cold thermal baths respectively, $Z$ is related to the
{\em efficiency} $\eta$ of converting  the heat  current $J_Q$
(between the baths) into the electric power $P$ which is generated
by attaching a thermoelectric element to an optimal Ohmic
impedance. Namely, in the linear regime:
\begin{equation} \label{eq:efficiency}
\eta = \frac{P}{J_Q} = \eta_\mathrm{Carnot} \cdot \frac{\sqrt{ZT +
1}-1}{\sqrt{ZT + 1} + 1} \ ,
\end{equation}
where $\eta_\mathrm{Carnot}=1-T_C/T_ H$ is the Carnot efficiency
and $T = (T_H  + T_C)/2$. Thus  a good  thermoelectric device  is
characterized  by a large value of the non-dimensional figure of
merit $ZT$. However, in spite of the fact that the second
principle of thermodynamics does not impose any restriction on the
value of $ZT$, all attempts to find high $ZT$ values (let us say
$ZT > 3$ at room temperature) have failed. The problem is that the
different transport  coefficients  $S$, $\sigma$ and $\kappa$ are
interdependent,  making the optimization of $ZT$ extremely
difficult. We believe that a better understanding of the possible
microscopic mechanisms \cite{linke} which determine the value of
$ZT$ may lead to a substantial improvement.

In a recent paper \cite{CasatiA}, a dynamical system approach to a
Lorenz gas type model has been used and an interesting mechanism to
reach high $ZT$ value has been discovered. More recently, a
one-dimensional, di-atomic disordered chain of hard-point elastic
particles has been considered \cite{CasatiB} and it has been found
that $ZT$ diverges to infinity with increasing the number of
particles inside the chain. This result suggests the possibility to
build a thermoelectric engine by connecting two heat baths with two
chains of different sizes.

In this paper we analyze such an engine and we show that, indeed, a
non zero circular current is established inside the system when the
stationary state is reached. Quite interestingly the efficiency of
such an engine appears to be unrelated to the figure of merit $ZT$
and it remains quite low in spite of the fact that $ZT$ becomes
larger and larger with increasing the system size. Indeed we
analytically show and numerically confirm that the efficiency of
such engine is determined by a new figure of merit $YT$. From one
hand the low value of $YT$ explains the poor efficiency of the
engine; on the other hand, it sheds new light on future directions
for increasing the efficiency of thermoelectric engines.

%%%%%%%%%%%%%%%%%%%%%%%%%%%%%%%%%%%%%%%%%%%%%%%%%%%%%%%%%%%%%%%%%%%%%%%%%%%%% Fig 1
\begin{figure}
\vspace{0.cm}
\includegraphics[width=.95\columnwidth,clip]{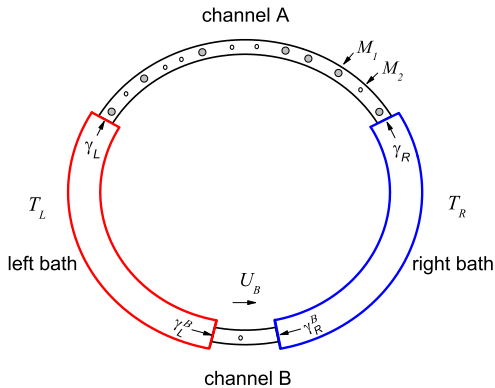}\vspace{-0.4cm}
\caption{Scheme of the thermoelectric engine studied in this
paper. It consists of two heat baths at different temperature
$T_L$ and $T_R$ and two channels of different lengths. See text
for the details.} \label{Fig1}
\end{figure}
%%%%%%%%%%%%%%%%%%%%%%%%%%%%%%%%%%%%%%%%%%%%%%%%%%%%%%%%%%%%%%%%%%%%%%%%%%%%% Fig 1

We consider a simple model of the engine (see Fig. \ref{Fig1}) which
consists of two heat baths and two connecting channels $A$ and $B$
of length $L_A$ and $L_B$ respectively. We assume that the whole
system does not exchange particles with the outside.  Therefore the total number of particles $N$ in the two
channels and in the two baths remains constant during the
simulation. The two heat baths are kept at different temperatures
$T_L$ and $T_R$, and have the same finite length $V$. Each channel
is a one-dimensional, di-atomic disordered chain, of hard-point
elastic particles with velocities $v_k$ and masses $m_k \in
\{M_1,M_2\}$ randomly distributed (we use convenient non-dimensional
units). The particles interact among each other through elastic
collisions only. A collision between two neighboring particles with
mass $m_k$ and $m_l$ causes a change of their velocities $v_k$ and
$v_l$ as
%$v'_k=\frac{m_k-m_l}{m_k+m_l}v_k+\frac{2m_l}{m_k+m_l}v_l$ and
%$v'_l=\frac{2m_k}{m_k+m_l}v_k-\frac{m_k-m_l}{m_k+m_l}v_l$.
$\Delta v_k={2m_l}(v_l-v_k)/({m_k+m_l})$ and $\Delta
v_l={2m_k}(v_k-v_l)/({m_k+m_l})$. An efficient algorithm has been developed which
performs correct chronological order of collisions and update particle's positions and velocities in 
$\sim\log N$ computer operations per collision.
We fix the length of the channel
B to be small and study how the properties of the engine vary as
the length of the channel A is increased. In particular, if
channel B is small enough then particles can pass trough it
without suffering any collisions.

The heat baths are modeled in the following way. Suppose $n_1$
particles of mass $M_1$ and $n_2$ particles of mass $M_2$ are
confined in a box of length $V$.  The particles collision rate with
one end of the box is given by
\begin{equation}\label{eq:gamma}
\gamma=\frac{\rho_1}{\rho}\gamma_1+\frac{\rho_2}{\rho}\gamma_2
\end{equation}
where $\gamma_i=\rho{\sqrt{\frac{T}{2\pi M_i}}}$, $\rho_i = n_i/V$
($i=1,2$), $\rho=n/V$ and $n = n_1 + n_2$.  Then the heat bath model
is straightforward: if one end of the box is opened, the particles
are emitted with the same rate $\gamma$, and the time interval $t$
between two consecutive emissions is a random variable which obeys the distribution
\begin{equation}\label{Prt}
Pr(t)=\frac{1}{t_0}e^{-\frac{t}{t_0}}
\end{equation}
with $t_0=1/\gamma$. The mass of the emitted particle is assigned
to be $M_i$ ($i=1,2$) randomly according to the probability
\begin{equation}\label{Pii}
\Pi_i=\frac{\rho_i\gamma_i}{\rho_1\gamma_1+\rho_2\gamma_2},
\end{equation}
%Eq.(\ref{Pii})
and its velocity is generated from the distribution
\begin{equation}\label{Piv}
P_i(v)=\frac{M_i|v|}{T}e^{-\frac{M_iv^2}{2T}}.
\end{equation}
As expected, when $M_1=M_2$, the heat bath model of identical
particles is recovered.

The emission rate from the left heat bath into channel A is
therefore $\gamma_L=\frac{\rho_{1,L}}{\rho_{L}}
\gamma_{1,L}+\frac{\rho_{2,L}}{\rho_{L}}\gamma_{2,L}$ where $\rho_L$
is the total particle number density at the left heat bath,
$\rho_{i,L}$ is that of particles with mass $M_i$, and
$\gamma_{i,L}=\rho_L{\sqrt{\frac{T_L}{2\pi M_i}}}$ ($i=1,2$).
%As $\rho_{L}$, $\rho_{L,1}$ and $\rho_{L,2}$ depend on time, so
%does $\gamma_L$.
Similarly one can write the expression for the emission rate
$\gamma_R$ from the right heat bath.

As to the channel B we assume that the emission rates are
proportional to $\gamma_L$ and $\gamma_R$:
\begin{equation}
\gamma_L^B=r\gamma_L,~~\gamma_R^B=r\gamma_R,
\end{equation}
where $r$ is an adjustable parameter. For $r=0$ channel B is in
fact closed hence there is no net particle current.

Conversely, whenever a particle from each channel arrives at the
border with the bath it is absorbed by the bath, so that a
stationary state is established after sufficiently long time. If
the stationary state is such that there is a net particle current
around the system, then one can use it to extract work. To this
end we insert inside the channel B an auxiliary potential $U_B$
which the particles have to climb thus performing useful work. Let
us first consider the case where the channel B is short enough
(i.e. $L_B=1$) so that the probability of particles collision
inside it is negligible. Suppose $T_L>T_R$ then the current runs
clockwise at $U_B=0$ due to the pressure balance between the two
heat baths. Thus a particle emitted from the right heat bath can
pass through the channel B only when its kinetic energy is larger
than $U_B$, otherwise it will return back. Therefore the
probability with which the emitted particle passes through the
channel B is
\begin{equation}
{\cal P}_{RL}=e^{-\frac{U_B}{T_R}}
\end{equation}
for both particles of masses $M_1$ and $M_2$. Hence the work
extracted per unit of time is
\begin{equation}
P=r(\gamma_R {\cal P}_{RL}-\gamma_L)U_B.
\end{equation}
The efficiency of the engine thus reads
\begin{equation}
\eta=\frac{P}{J_{Q,A}+J_{Q,B}}
\end{equation}
where $J_Q^A$, and $J_Q^B$, are the net thermal energy flows from the
left heat bath in a unit time into the channel A, and B,
respectively, which can be measured by numerical simulations. Notice that,
in particular, $J_Q^B$ can be expressed analytically through
\begin{equation}
J_{Q,B}=r(T_L \gamma_L-{\cal P}_{RL} T_R \gamma_R).
\end{equation}

%%%%%%%%%%%%%%%%%%%%%%%%%%%%%%%%%%%%%%%%%%%%%%%%%%%%%%%%%%%%%%%%%%%%%%%%%%%%% Fig 2
\begin{figure}
\vspace{0.cm}
\includegraphics[width=.95\columnwidth,clip]{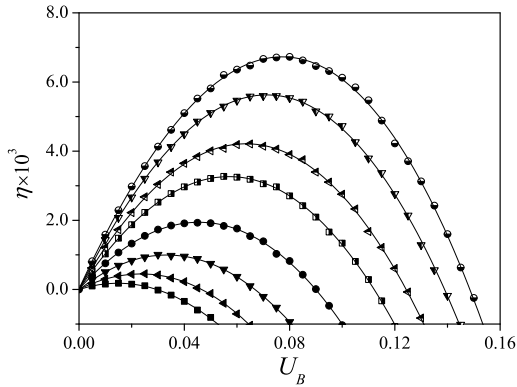}\vspace{-0.4cm}
\caption{Efficiency  of the engine model illustrated in Fig.
(\ref{Fig1}) as a function of the potential $U_B$ for $L_A$=4, 8,
16, 32, 64, 100, 200 and 400 (from below to up). $T_L=1.2$,
$T_r=0.8$, and $r=1$. The length of channel B is unity.}
\label{Fig2}
\end{figure}
%%%%%%%%%%%%%%%%%%%%%%%%%%%%%%%%%%%%%%%%%%%%%%%%%%%%%%%%%%%%%%%%%%%%%%%%%%%%% Fig 2

As it is seen from Fig. \ref{Fig2} our engine works.  Here we plot
the numerically computed efficiency $\eta$ against the potential $U_B$ for various
lengths of the channel A. The solid curves are polynomial (3-order)
fittings of the numerical data.

%based on which the
%maximum efficiency $\eta_{max}$ at certain value of potential
%$U_B=U_{B,max}$ for each $L_A$ can be accurately evaluated.
In this calculation the length of channel B is set to unity and the
length of both heat baths is very large, namely $V=10^3$. The total
number of particles in the system is set to be $N=2V+L_A+ L_B$ such
that the overall averaged particle density is unity. The two types
of molecules are set to be equal in number and have masses $M_1=1$
and $M_2\approx 0.618$ respectively. In addition, for a given length
$L_A$ of the channel A, there is an optimal value of $U_B$
%(denoted by $U_{B,\mathrm{max}}$)
such that the efficiency is maximized. Denoting the maximum
efficiency by $\eta_\mathrm{max}$, it is interesting to compute the
dependence of $\eta_\mathrm{max}$ on $L_A$ and compare it with theoretical
expectations.
%and as suggested by Fig. \ref{Fig2}.

%%%%%%%%%%%%%%%%%%%%%%%%%%%%%%%%%%%%%%%%%%%%%%%%%%%%%%%%%%%%%%%%%%%%%%%%%%%%% Fig 3
\begin{figure}
\vspace{0.cm}
\includegraphics[width=.95\columnwidth,clip]{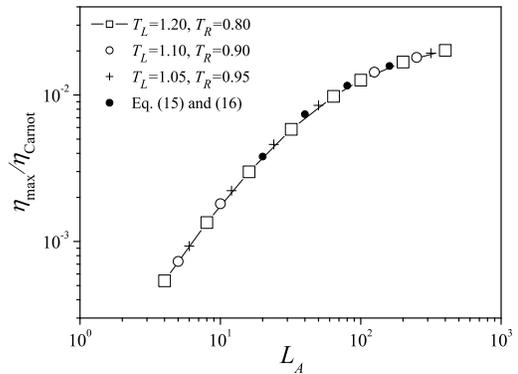}\vspace{-0.4cm}
\caption{The maximum efficiency  of the engine for different
temperature gradients. Here $r$=1 and $L_B=1$. The solid dots are
obtained from Eq. (\ref{etamax}) and Eq. (\ref{YT}), where the
parameters $\kappa$, $\sigma$, $S$ for both channels are numerically
obtained as in Fig. (\ref{Fig4}). As it is seen there is perfect
agreement with the directly, numerically computed, efficiency.
% $M_1=1, M_2\approx 0.618, T=1,
%\mu=1$ and $2\rho_1=2\rho_2=\rho=1$.
} \label{Fig3}
\end{figure}
%%%%%%%%%%%%%%%%%%%%%%%%%%%%%%%%%%%%%%%%%%%%%%%%%%%%%%%%%%%%%%%%%%%%%%%%%%%%% Fig 3

Extensive and accurate numerical computations, summarized in Fig.
\ref{Fig3}, show that even though $\eta_{\mathrm{max}}$
%The maximum efficiency $\eta_{max}$ for each $L_A$ value in Fig.
%(\ref{Fig2}) can be accurately evaluated based on the
%corresponding polynomial fitting. The results are summarized in
%Fig. \ref{Fig3}, together with the repeating calculation results
%with different temperature pairs ($T_L, T_R$).
increases with $L_A$, i.e. with the average number of molecules in
the channel A, the increasing rate slows down very fast. Also it
seems hard to find a simple fitting for the dependence of
$\eta_{\mathrm{max}}$ on $L_A$.

We should notice that $\eta_{\mathrm{max}}$ depends also on $T_L$
and $T_R$. However if $\eta_{\mathrm{max}}$ is rescaled to the
Carnot efficiency, its behavior with $L_A$ is the same for all
($T_L$, $T_R$) pairs, given that the temperature difference is
small enough. This can be seen clearly in Fig. \ref{Fig3}.

{\it Remarks}: We have also investigated carefully several variants
of the engine model. First, we have studied the dependence of
$\eta_{\mathrm{max}}$ on $r$. We have found that indeed
$\eta_{\mathrm{max}}$ can be slightly improved by adjusting $r$
(e.g. $\sim 30\%$ as compared to Fig. \ref{Fig3}), however the
efficiency remains very low. Then we have checked whether adding a
potential $U_A$ against the current in the channel A can improve
efficiency. In this case a particle in the channel A would undergo a
parabolic motion between two consecutive collisions with its
neighbors. This variant does not increase the maximum efficiency
$\eta_{\mathrm{max}}$ either which has been found to depend only on
the sum $U_A+U_B$ rather than on $U_A$ and $U_B$ separately. Also a
longer length of channel B has been considered but it turns out that
for a given $L_A+L_B$, the efficiency reaches its highest value when
$L_B\leq 1$. To summarize, the various modifications to the engine
model we have considered do not improve the efficiency
significantly.

%%%%%%%%%%%%%%%%%%%%%%%%%%%%%%%%%%%%%%%%%%%%%%%%%%%%%%%%%%%%%%%%%%%%%%%%%%%%% Fig 4
\begin{figure}
\vspace{-.4cm}
\includegraphics[width=.95\columnwidth,clip]{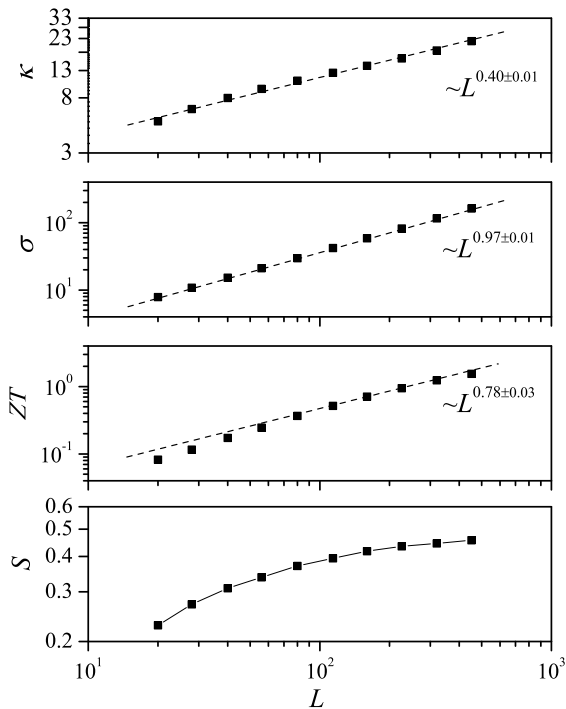}\vspace{-2.cm}
\caption{Dependence on the channel length of various parameters of
the 1D mixed gas. From these results it is clear why when $ZT$ can
be very large for a long channel, the new merit $YT$ that governs
the efficiency of the engine model (Fig. \ref{Fig1}) takes a lower
value instead.} \label{Fig4}
\end{figure}
%%%%%%%%%%%%%%%%%%%%%%%%%%%%%%%%%%%%%%%%%%%%%%%%%%%%%%%%%%%%%%%%%%%%%%%%%%%%% Fig 4

Let us now explain why our engine works at a rather low efficiency
in spite of the fact that $ZT$ goes to infinity. In order to derive
a theoretical expression for the efficiency we use thermodynamic
linear response relations for the heat currents $J_{Q,A}$,
$J_{Q,B}$, and for the particle currents $J_{\rho,A}$, $J_{\rho,B}$,
in the channels A and B, respectively, namely
\begin{eqnarray}
J_{Q,A}&=&-\kappa_A'\Delta T/L_A -T\sigma_A S_A (\Delta \mu/L_A +U_A)
\nonumber\\
J_{\rho,A}&=&-\sigma_A S_A\Delta T/L_A-\sigma_A (\Delta\mu/L_A +U_A)~~~
\nonumber\\
J_{Q,B}&=&-\kappa_B'\Delta T/L_B-T\sigma_B S_B (\Delta\mu/L_B-U_B)~~
\label{currents}\\
J_{\rho,B}&=&-\sigma_B S_B\Delta T/L_B-\sigma_B (\Delta\mu/L_B-U_B)~~~~~
\nonumber
\end{eqnarray}
where $\kappa_A'=\kappa_A+T\sigma_A S_A^2$,
$\kappa_B'=\kappa_B+T\sigma_B S_B^2$, $\mu$ is the chemical
potential, and $\Delta T = T_R- T_L$, $\Delta\mu = \mu_R - \mu_L$.
In addition, in the stationary state we must also have
\begin{equation}
J_{\rho,A}+J_{\rho,B}=0. \label{sumc}
\end{equation}

The relations given by Eq. (\ref{currents}) and (\ref{sumc})
represent five linear equations in the five unknowns $J_{Q,A}$,
$J_{Q,B}$, $J_{\rho,A}$, $J_{\rho,B}$, and $\Delta\mu$, whereas the
other parameters, such as transport or Onsager coefficients are
considered to be known.
From the solutions of the above equations we compute the efficiency
$\eta$ as a function of the two bias potentials:
\begin{equation}
\eta(U_A,U_B)=\frac{P}{J_Q}=\frac{J_{\rho,A}U_A-J_{\rho,B}U_B}
{J_{Q,A}+J_{Q,B}}
\end{equation}
Since $\eta(U_A,U_B)$ is a
quadratic function, we can easily maximize it by computing the
optimal values of $U_A,U_B$. The final result can be expressed in a
simple form as
\begin{equation}\label{etamax}
\eta_{\mathrm{max}}/\eta_{\mathrm{Carnot}}=
1-2\sqrt{(YT)^{-2}+(YT)^{-1}}+2(YT)^{-1}
\end{equation}
with a new figure of merit
\begin{equation}\label{YT}
YT=\frac{(\sigma_A/L_A)(\sigma_B/L_B)}{\sigma_A/L_A+\sigma_B/L_B}\cdot\frac{(S_A
-S_B)^2}{\kappa_A/L_A+\kappa_B/L_B}\cdot T
\end{equation}
From Eq. (\ref{etamax}) it can be seen that efficiency is
given by this new figure of merit exclusively and it has nothing
to do with the merit $ZT$. Clearly, as $YT\to 0$ ($YT\to \infty$)
we have $\eta_{\rm max}\to 0$ ($\eta_{\rm max}\to \eta_{\rm
Carnot}$).

In order to understand why our engine model has a low efficiency, we
investigate carefully the dependence of the parameters $\sigma$,
$\kappa$ and $S$ on the channel length. These quantities can be
obtained from the Onsager coefficients \cite{CasatiA} which in turn
are calculated by measuring the particles current $J_\rho$, and the
energy current $J_u$, when the stationary state of the system is
reached through
\begin{eqnarray}
J_u=L_{uu}\Delta\beta/L-L_{u\rho}\Delta\alpha/L
\nonumber \\
J_\rho=L_{\rho
u}\Delta\beta/L-L_{\rho\rho}\Delta\alpha/L
\end{eqnarray}
%With the Onsager
%coefficients, it is straightforward to obtain the ZT merit with
%\begin{equation}
%ZT=\frac{L_{u\rho}L_{\rho u}}{\det L}.
%\end{equation}
with $\alpha\equiv \mu/T$, $\beta\equiv 1/T$ and $\mu$ the chemical
potential:
\begin{equation}\label{mu}
\mu/T=(\rho_1 \mathrm{ln} \rho_1+\rho_2
\mathrm{ln} \rho_2)/\rho - \mathrm{ln}\sqrt{T} + {\rm const}.
\end{equation}

In Fig. \ref{Fig4} we present the numerical results for a 1D mixed
gas chain with $M_1=1, M_2\approx 0.618, T=1$ and
$2\rho_1=2\rho_2=\rho=1$. The constant in Eq. (\ref{mu}) is set to
be $\mathrm{ln}2+1$ such that $\mu=1$. The results clearly show
that, as the channel length is increased, while the ratio
$\sigma/\kappa$ increases as a power law, the Seebeck coefficient
$S$ undergoes a slower and slower increase. As a result, while $ZT$
increases indefinitely with the channel length $L$, the merit $YT$
remains small.

In summary, for the engine model (Fig. \ref{Fig1}) based on a one
dimensional mixed gas, the efficiency is governed by a new figure of
merit. Based on its relation to the thermoelectric parameters (Eq.
(\ref{YT})), it can be expected that an efficient mixed gas type
engine should be such that its Seebeck coefficient changes fast when
the channel length is changed. Such a mixed gas might be that
consisting of three, or more, types of molecules. We believe that
our method of analysis of thermoelectric or thermochemical heat
engines should be applicable to a wide range of models which consist
of two transport channels between a pair of baths.
% which has not been explored in previous studies.

JW acknowledges support from the Defense Science and Technology
Agency (DSTA) of Singapore under agreement of POD0613356, and TP
acknowledges Grants P1-0044 and J1-7347 of the Slovenian
research agency.


\begin{thebibliography}{99}

\bibitem{rev} S. Lepri, R. Livi, and A. Politi, Phys. Rep.
{\bf 273}, 1 (2003); A. Dhar,  Adv. Phys. {\bf 57}, 457 (2008); F.
Bonetto, J. L. Lebowitz and L. Ray Bellet, in ``Mathematical Physics
2000", A Fokas, A Grigoryan, T Kibble, B Zegarlinski (eds)(Imperial
College Press, London, 2000) (pg. 128-150); G. Casati, J. Ford, F.
Vivaldi and W. M. Visscher, Phys. Rev. Lett. {\bf 52}, 1861 (1984).

\bibitem{mahan} G. Mahan, B. Sales, J. Sharp, Phys. Today {\bf 50}, 42 (March 1997).

\bibitem{majumdar} A. Majumdar, Science {\bf 303}, 777 (2004).

\bibitem{dresselhaus} M. S. Dresselhaus {\em et al}, Adv. Mater. {\bf 19}, 1043-1053 (2007).

\bibitem{linke} T. E. Humphrey and H. Linke,  Phys. Rev. Lett. {\bf  94}, 096601 (2005).
\bibitem{CasatiA} G. Casati, C. Mejia-Monasterio, and T. Prosen,
         Phys. Rev. Lett. {\bf 101}, 016601 (2008)
\bibitem{CasatiB} G. Casati, L. Wang, and T. Prosen,
         {\it A One-Dimensional Hard-point gas and the thermoelectric efficiency}, preprint (2008)
\end{thebibliography}
\end{document}